\documentstyle[prl,aps,twocolumn]{revtex}
\include{psfig}

\begin{document}
\title{Micro-mechanical charge transfer mechanism in 
soft Coulomb blockade nanostructures}

\title{Shuttle mechanism for charge transfer in Coulomb
blockade nanostructures}
\author{
L. Y. Gorelik,$^{1,2}$ A. Isacsson,$^{1}$ M. V. Voinova,$^{1,3}$
B. Kasemo,$^{1}$ R. I. Shekhter,$^{1}$ and M. Jonson$^{1}$
}
\address{ $^{(1)}$Department of Applied Physics, Chalmers University of 
Technology and G{\"o}teborg University, S-412 96 G{\"o}teborg, Sweden\\
$^{(2)}$ B. Verkin Institute for Low Temperature Physics and Engineering,
310164 Kharkov, Ukraine\\
$^{(3)}$ Kharkov State University, 310077 Kharkov, Ukraine\\
}


\maketitle
\begin{abstract}
Room-temperature Coulomb blockade of charge transport through composite
nanostructures 
containing organic inter-links has recently been observed. A pronounced 
charging effect in combination with the softness of the molecular links 
implies that charge transfer gives rise to a significant deformation of 
these structures. For a simple model system containing one nanoscale metallic 
cluster connected by molecular links to two bulk metallic electrodes we show
that self-excitation  of periodic cluster oscillations in conjunction 
with sequential processes of cluster charging and decharging appears for 
a sufficiently large bias voltage. This new `electron shuttle' mechanism 
of discrete charge transfer gives rise to a current through the 
nanostructure, which is proportional to the cluster vibration frequency.

\vspace{3mm}
\vspace{3mm}
\end{abstract}

The term Coulomb blockade refers to a supression of the tunneling current 
through metallic grains embedded in a dielectric matrix. The origin of this 
phenomenon lies in the fundamental quantization of charge in units of $e$ and 
it occurs in systems with small quantum charge fluctuations. 
As a result the redistribution of grain charges necessarily associated
with a current flow can only be made in quantized units of $e$. 
The correspondingly quantized electrostatic charging energy, which for small 
nanoscale grains can be large compared to other relevant energies related
to temperature and bias voltage, tends to block the current.
A number of different Coulomb blockade-based phenomena have been discovered
--- typically at very low temperatures --- 
over the last decade or so \cite{one,L}, largely due to the developments
within nanotechnology. 

In this Letter we discuss 
a new type of composite mesoscopic structure demonstrating
Coulomb blockade behavior at room temperatures \cite{two,three,four}.
The crucial aspect of these structures from the point of view of
our work is that they contain metallic grains or molecular
 clusters with a 
typical size of 1-5 nm that can vibrate; their positions are not 
necessarily fixed.
This is because the dielectric material surrounding them is elastic and
consists of mechanically soft organic molecules. 
These molecular inter-links have elastic moduli which
 are typically two or three orders
of magnitude smaller than those of ordinary solids \cite{Mm}.
Their
ohmic resistance $R$ is high and of order $10^7$ - $10^8$ ohm, while
at the same time they are
extremely small --- a few nanometers in size.
\begin{figure}[!t]
  \begin{center}
    \leavevmode
    \centerline{\psfig{figure=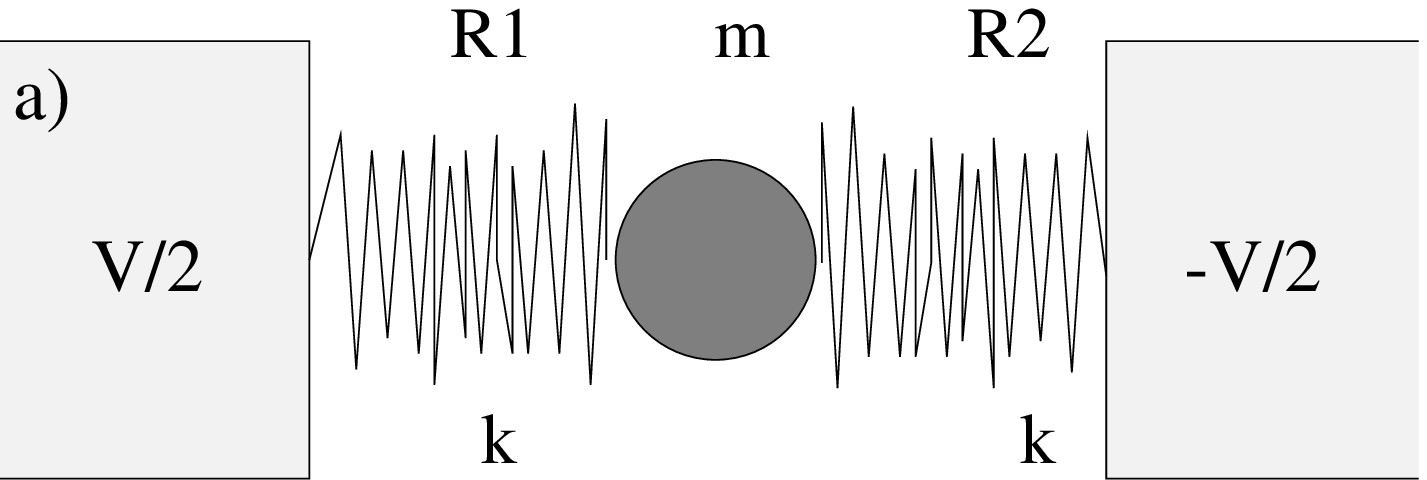,width=8cm}}
    \centerline{\psfig{figure=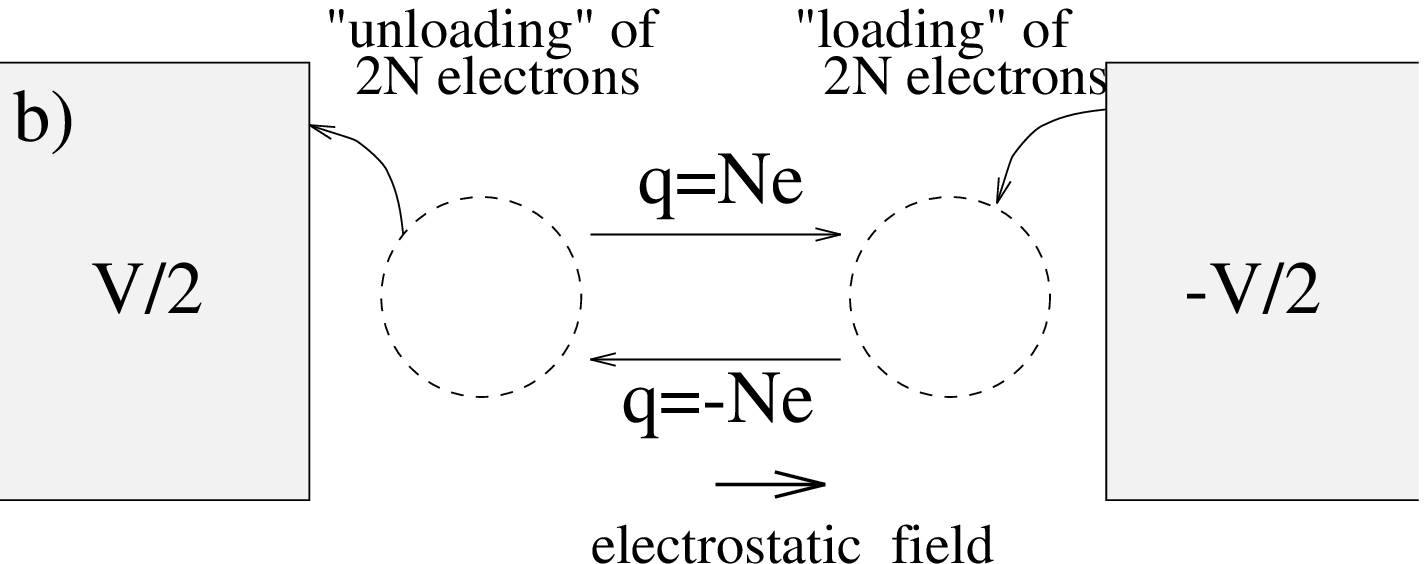,width=8cm}}
    \caption{(a) Simple model of a soft Coulomb blockade system in which a metallic
      grain (center) is linked to two electrodes by elastically deformable organic
      molecular links. (b) Dynamic instabilities occur since in the
      presence of a sufficiently large
      bias voltage $V$ the grain is accelerated by the same
      electrostatic force towards first one, then the other electrode. A cyclic
      change in direction is caused by the repeated ``loading'' of  electrons
      near the negatively biased electrode and the subsequent ``unloading'' of the
      same at the positively biased electrode. 
      As a result the sign of the net grain charge
      alternates leading to an oscillatory grain motion and a novel ``electron
      shuttle" mechanism for charge transport.
      }
    \label{fig:one}
  \end{center}
\end{figure}
A large Coulomb blockade effect in
combination with the softness of the dielectric medium implies that 
charge transfer may give rise to a significant deformation of these 
structures as they
respond to the electric field associated with a bias voltage.
 We have found that self-excitation of mechanical 
grain vibrations accompanied by barrier deformations is possible for a 
sufficiently large bias voltage. As we show below this gives rise to a
novel `shuttle mechanism' for electron transport. The point is that a
grain (the shuttle) oscillates between two turning points. One of them is
located near a positively- and the other near a
negatively biased electrode. Due to the Coulomb blockade phenomenon
an integer number of electrons are loaded onto the grain close to one 
turning point and the same number of electrons are unloaded close to the
other as illustrated in Fig.~1. The result is that in each 
cycle the shuttle moves
a discrete number of electrons from one electrode to the other. It follows
that the current is proportional to the mechanical vibration frequency 
of the grain. 

One can estimate the typical
frequency $\omega$ of such vibrations using elastic moduli for the relevant
organic molecules. For a metallic cluster a few nanometers in diameter and
a molecular length of 1~nm the characteristic vibration frequency
is of the order of $10^{10} - 10^{11} {\rm s}^{-1}$. 
Hence it can be of the order of or higher than the typical charge fluctuation
frequency 
$\nu = 1/RC $ (here $C$ is the capacitance of the metallic cluster which
for room temperature Coulomb blockade systems is of the order
$10^{-18} -10^{-19}$ F). This means that charge fluctuations and
cluster vibrations become strongly coupled in such systems.

For illustration we consider the simplest possible interesting system. It 
consists of one small metallic grain 
connected by elastic molecular links to two bulk leads on either side,
as shown in Fig.~1a.
The tunnel junctions between the leads and the grain are modeled
by 
tunneling 
resistances $R_1(x)$, $R_2(x)$ which are assumed to be exponential
functions of the grain coordinate $x$.
In order to avoid unimportant technical complications
we study the symmetric case for which $R_{1,2} = R e^{\pm x/\lambda}$.
When the position of the grain is fixed, the electrical potential
of the grain and its charge $q$ are given by current balance
between the grain and the leads \cite{one}. As a consequence, at a given
bias voltage $V$ the charge $q$ is completely controlled by the ratio
$R_1(x)/R_2(x)$. 
In addition the bias voltage generates an electrostatic
field ${\cal E}= \alpha V$  in the space 
between the leads \cite{C} and hence a 
charged grain will
be subjected to an electrostatic force $F_q = \alpha Vq$.

The central point of our considerations is that
the grain --- because of the ``softness'' of the organic molecular links
connecting 
it to the leads --- may move and change its position.
The grain motion disturbs the current balance and as a result 
the grain charge will vary in time. This variation
affects the work $W = \alpha V\int \dot{x}q(t)dt$ 
performed on the grain during, say, one period of its oscillatory motion.
It is significant that this work is nonzero and positive, i.e., 
the electrostatic
force, on the average, accelerates the grain. The nature of
this acceleration is best understood by considering a
grain oscillating with a large amplitude $A > \lambda$. 
In this case the charge fluctuations between the grain and the most distant
electrode is 
exponentially suppressed when the displacement of the grain is maximal. 
As a result 
the grain at such a turning point gains extra charge from the nearby 
electrode, as shown in Fig.~1b. The added charge 
is positive at the turning
point near the positively charged electrode and negative at the turning 
point close to the negatively charged one. The sign change of the charge
takes place mainly in the immediate vicinity of the
electrodes while along the major part of the path between them 
the charge on the grain is fixed and determined by the most
recently encountered electrode. Therefore the grain acts as a shuttle that
carries positive charge on its way from the positive to the negative electrode
and negative charge on its return trip. 
The electrostatic force is hence at all times 
directed along the line of motion
and thus accelerates the grain.
This statement is true also for small amplitude oscillations. The
result is an electro-mechanical instability
in the sense that even if the grain is initially at rest a
sufficiently strong
electrostatic field will cause the grain to start vibrating.
\begin{figure}[!t]
  \begin{center}
    \leavevmode
    \centerline{\psfig{figure=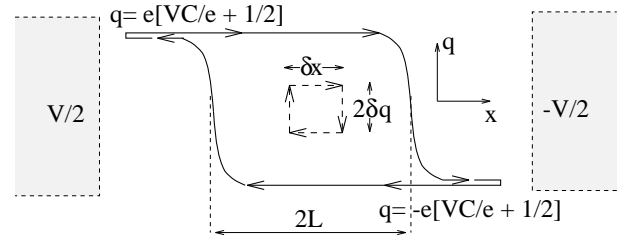,width=8cm}}
    \vspace{0.3cm}
    \caption{Charge response to a cyclic grain motion. The dashed arrows 
describe an imagined, particularily simple trajectory in the charge-position
plane that allows the work done on the grain by the electrostatic field
to be easily calculated and shown to be positive, hence leading to an 
instability: For times $t < -\tau$ when the grain is at rest at
$x = -\delta x/2$ 
the charge exchange with the positively biased electrode dominates and
the grain is positively charged $q = \delta q$; at the instant $t=-\tau$
it instantaneously moves to the point $x = +\delta x/2$ where the
charge relaxes to a new equlibrium value $q =- \delta q$; then the
negatively charged grain instantaneously moves back at $t=+\tau$.
During this cyclic process the electrostatic force acts only
along the direction of the grain displacement.
The solid line describes (realistically)
the charge response at large oscillation amplitudes.
At $|x|>L$ the charge exchange with the far lead is
exponentially suppressed and the charge on the grain takes the
fixed quantized value corresponding to thermal equilibrium with the
nearest lead. As a result no net energy is pumped into the system.
}
    \label{fig:two}
  \end{center}
\end{figure}
To be more precise one can show that for
small deviations from equilibrium ($x=0$, $q = 0$) and
provided $q(t)$ is
defined as the linear response to the grain displacement,
$q(t) = \int \chi (t-t')x(t')dt'$,
the work done on the grain is positive.
To prove this one may consider the simple closed 
trajectory in $x$,~$q$-space shown with dashed arrows in Fig.~2.
For this trajectory it follows
trivially that the work $W$ is positive. Since this result can be shown \cite{Sign}
to be independent of the particular trajectory considered, 
this simplifying choice does not imply any restriction on its validity.
The positive value of $W$ at small vibration amplitudes
implies that self-excitation of mechanical vibrations will occur
when a finite voltage
bias $V$ is applied to our system.

In real systems a certain amount $Q$ of energy is dissipated
due to mechanical damping forces which always exist. 
In order to get to the self-excitation regime more energy must
be pumped into the system from the electrostatic field 
than can be dissipated; $W$ must exceed $Q$.
Since the electrostatic force increases with the 
bias voltage this condition can be fullfiled if $V$ exceeds some
critical value $V_c$. 
If the electrostatic and damping forces are much smaller 
than the elastic restoring force self-excitation of vibrations 
with a frequency equal to the eigenfrequency 
of elastic oscillations arise. 
In this case $V_c$ can be implicitly defined
by the relation $\omega\gamma =  \alpha V_c {\rm Im}\chi(\omega)$, 

where $\omega\gamma$ is the imaginary part of the complex dynamic 
modulus.
In the general case, when the charge response is determined by
Coulomb-blockade phenomena, $\chi$ is an increasing but rather
complicated function of $V$ and there is no way to solve
for $V_c$ analytically. 
However, one can establish the value of $V_c$ with an uncertainty 
given by the Coulomb blockade voltage $V_e = e/C$:
%

Above the threshold voltage  the oscillation amplitude will
increase exponentially until a balance between 
dissipated and absorbed energy is achieved and the system reaches a stable 
self-oscillating regime\cite{ch}. The amplitude $A$ of the
self-oscillations will therefore be determined by the criterion 
 $W(A) = Q(A)$.
Let us now consider large amplitude oscillations.
Then, due to the exponentially strong dependence of
the tunnel resistances on the grain position,
no net energy is absorbed by the system outside the region
$|x| < L = \ln{A\omega/\lambda \nu} < A$ (see Fig.~2).
Therefore, the amount of absorbed energy in this limit does not depend
strongly on the amplitude $A$ \cite{La}.
This results in a saturation of the energy that can be pumped into
the system, which is of
the order $W \sim 4\alpha V q_{eq}L$. 
Here $q_{eq}$ is the magnitude of the charge that obtains if
the grain is in thermal equilibrium with one lead.
In the Coulomb blockade regime, at temperatures $k_BT \ll e^2/C$,
the value of $q_{eq}$ is quantized in units of $e$:
$q_{eq} = Ne = \left[ VC/e + 1/2 \right]e$.
To estimate the dissipated energy  we take
$Q(A) = \pi \omega \gamma A^2$, which corresponds to the simple 
phenomenological 
damping force $F_d = -\gamma \dot{x}$.
This approach leads to the following expression for the amplitude of
self-oscillation when $V\gg V_c$:
$A \sim \lambda \Omega VC/e(\gamma \omega)^{1/2}$ ($\Omega$ is defined 
in Fig.~3). One can see 
from this expression that the self-oscillation grows in amplitude with
increasing bias voltage.

In the fully developed self-oscillating regime  
the oscillating grain, sequentially moving electrons from one lead
to the other, provides a new `shuttle mechanism' for charge as shown 
in Fig.~1b. In each cycle $2N$ electrons are 
transferred, so the average contribution to the current from this
shuttle mechanism is
\begin{equation}
\label{Ieq}
I=2eNf \, , \quad N=\left[\frac{CV}{e}+\frac{1}{2}\right] \, ,
\end{equation}
where $f\equiv\omega/2\pi$ is the self-oscillation frequency. This current
does not depend on the tunneling rate $\nu$. The reason is that when the 
charge jumps
to or from a lead, the grain is so close that the tunneling
rate is large compared to the elastic vibration frequency. Hence the
shuttle frequency
--- not the tunneling rate --- provides the `bottle neck' for
this process.

We emphasize that the current due to this shuttle mechanism can be
substantially larger than the conventional current via a fixed grain.
This is the case when $\omega\gg \nu$. Our analysis shows that the
electromechanical instability discussed above has
dramatic consequences for the current-voltage characteristics of single
electron transistor configurations as shown in Fig.~3.
Even for a symmetric double junction, where
no Coulomb staircase appears in conventional designs, we predict that
the shuttle mechanism for charge transport manifests itself as a current
jump at $V=V_c$ and as a Coulomb staircase as the voltage is 
further increased.

To support the qualitative arguments given above we have performed
analytical and numerical analyses based on the simultaneous solution
of Newton's equation,
\begin{equation}\label{N}
 m \ddot{x}=-kx-{\gamma}\dot{x}+\alpha Vq, \; q=e\sum_n nP_n,
\end{equation}
for the motion of a grain of mass $m$ and a Master 
equation for the charge distribution function
$P_n \equiv {\rm Tr}\{\delta(n-\hat{n})\hat{\rho}\}$
($\hat{n}$ is the electron number operator and
$\hat{\rho}$ is the density matrix of electrons), 
\begin{eqnarray}\label{M}
  \label{tunrat}
  \frac{2}{\nu}&&\dot{P_n} = e^{-x/\lambda}\Gamma({n-1},{n})
P_{n-1}+e^{x/\lambda}\Gamma({n+1},{n})P_{n+1}
\nonumber \\
 &&-e^{-x/\lambda}\Gamma({n},{n+1})P_{n}-
e^{x/\lambda}\Gamma({n},{n-1}) P_n \, .
\end{eqnarray}
The quantities $P_n$ completely describe
the ``charge" state of the grain in the Coulomb blockade 
regime and hence the average current through the system
\begin{eqnarray}
  I=\frac{\nu}{2T}\int_0^T{e^{-x/\lambda}\sum_n{\Gamma(n,n+1)P_n}dt}
\end{eqnarray}
The dimensionless tunneling rates 
$\Gamma({n\mp 1},{n})$ in Eq.~(\ref{tunrat}) are given by
$$ \Gamma(n\mp 1,{n})=\left({CV\over{e}}\mp
n + \frac{1}{2}\right)\theta\left(\frac{CV}{e}\mp
n +{1\over{2}}\right).
$$
and numerically obtained $I-V$ curves for different amounts of 
damping are shown in Fig~3. 

A more precise (and time consuming) calculation along the line $l$ sketched 
in Fig.~3 is shown in Fig.~4. Rather than solving the Master Eq.~(\ref{M})
we here calculate the current utilizing a Monte Carlo type algorithm.
The nonmonotonic behaviour of the current along this line is due to 
competition between the two charge transfer mechanisms present in the system,
the ordinary tunnel current and the mechanically mediated current 
$I_{mech}(x_0,t)=\delta(x(t)-x_0)\dot{x}(t)q(t)$  through some cross section
at $x_0$. We define the shuttle current as the time average mechanical
current through the plane located at $x_0=0$. This current together
with the tunnel current for the same cross section is shown in Fig.~4. 
As the damping in the system is reduced the oscillation amplitude 
grows and the shuttle current is enhanced while the ordinary 
tunneling current is suppressed. In the limit of low damping this
leads to a quantization of the total current in terms of $2ef$.  
\begin{figure}[!t]
  \begin{center}
    \leavevmode
    \centerline{\psfig{figure=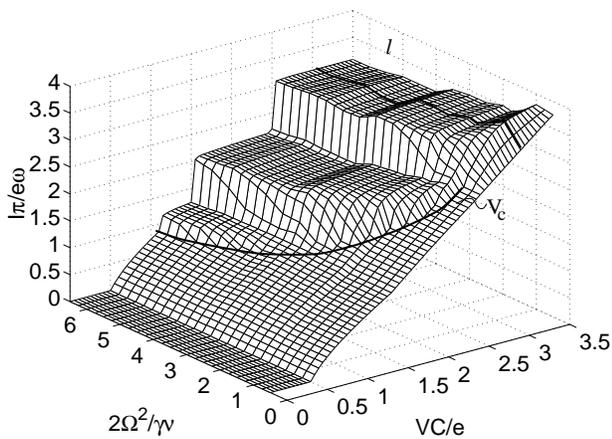,width=8cm}}
    \vspace{0.3cm}
    \caption{
      Current due to the proposed shuttle mechanism
      through the composite Coulomb blockade system of Fig.~1. The current is
      normalized to the eigenfrequency $\omega$ of elastic grain vibrations
      and plotted as a
      function of normalized bias voltage $V$ and inverse damping rate $\gamma^{-1}$
      ($\Omega = \sqrt{\alpha e^2/\lambda C}$). With infinite damping no grain 
      oscillations occur and no Coulomb staircase
      can be seen. The critical voltage $V_c$ required for the grain to start 
      vibrating is indicated by a line.
      }
    \label{fig:three}
  \end{center}
\end{figure}

\begin{figure}[!t]
  \begin{center}
    \leavevmode
    \centerline{\psfig{figure=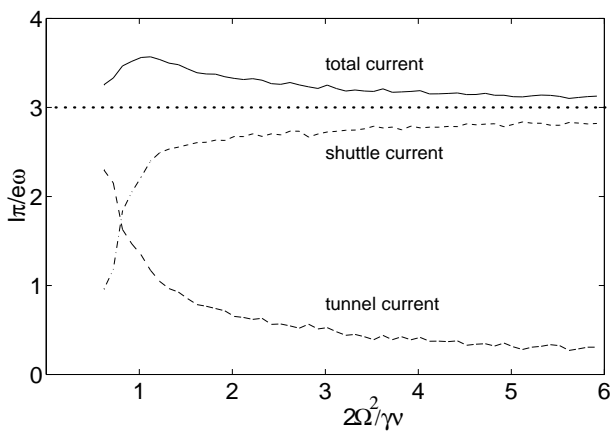,width=8cm}}
    \vspace{0.3cm}
    \caption{Cross section along the line $l$ in Fig.~3. The total time averaged 
      current consist of two parts, the shuttle current and the tunneling current. 
      The time averaged shuttle current is the mechanically transfered  
      current through the center of the system $<\delta(x(t))\dot{x}(t)q(t)>$, 
      the remaining part comes from ordinary tunneling. As the inverse 
      damping $\gamma^{-1}$ increase the shuttle current approaches the quantized 
      value $I\pi/e\omega=3$. The tunnel current is proportional to the fraction 
      of the oscillation period spent in the middle region, $|x|<\lambda$. This fraction 
      is inversely proportional to the oscillation amplitude and hence the tunnel 
      current decreases as $\gamma^{-1}$ increases. The fine structure in the 
      results is due to numerical noise.   
      }
    \label{fig:four}
  \end{center}
\end{figure}

In conclusion we have performed a numerical analysis of the charge transport 
by the shuttle mechanism through the
soft Coulomb blockade system shown in Fig.~1. The following results
summarize the discussion above: (i) a dynamical instability exists
above a critical bias voltage $V_c$; (ii) a limit cycle in the position-charge
plane exists for the grain (shuttle) motion above $V_c$ leading to a novel
`electron shuttle' mechanism for charge transport;
(iii) The current-voltage curve has a step-like structure, a type of Coulomb
staircase, even for a symmetric system. 
Our numerical analysis was made for 
a symmetric case for which $R_1(0) = R_2(0)$.
However all phenomena discussed above --- with small 
quantitative corrections --- still appear in the
asymmetric case when  $R_1(0) \neq  R_2(0)$, which is of great importance
for experiments. 
In a strongly asymmetric situation when  $R_1(0) \gg R_2(0)$ the 
electromechanical instability still exist but instead of
one threshold voltage $V_c$ a set of instability ``windows" 
associated with the $I-V$ steps \cite{GS} will appear.
This work has been supported by the Swedish TFR, KVA, and NFR.


\begin{thebibliography}{30}
\bibitem{C}
In reality the capacitances of the system and hence its
electrostatic energy $E(x)$ depend on the grain position.
By expanding $E(x)$ for fixed $q$ and $V$ and retaining only the
linear term one gets the position independent effective electrostatic
field used here.
\bibitem{ch}
Our numerical simulation show that instability develops 
into a self-oscillatory regime corresponding
to a limiting cycle rather than to a more complicated 
attractor or even chaotic motion.
\bibitem{La}
A more sophisticated treatment shows that 
$W \propto \ln{A}$ at large amplitude.
\end{thebibliography}
\end{document}